\documentclass[aps,prl,amsmath,amssymb,floatfix,twocolumn,amsmath,superscriptaddress,twocolumn,nofootinbib,tighten,letterpaper]{revtex4-1}
\usepackage{multirow}
\usepackage{bbold}
\usepackage{subfigure}
\usepackage{color}
\usepackage{mathrsfs}
\usepackage{hyperref}
\usepackage[normalem]{ulem}
\usepackage{bm}

\usepackage[utf8]{inputenc}

\usepackage{amsfonts, relsize, color}
\usepackage{graphics}
\usepackage{graphicx}
\usepackage{subfigure}
\usepackage{hyperref}
\usepackage{color}

\begin{document}

\title{\bf Shear viscosity as a probe of nodal topology}

\author{Marianne Moore}
\affiliation{Max-Planck-Institut f\"{u}r Physik komplexer Systeme, N\"{o}thnitzer Str. 38, 01187 Dresden, Germany}
\affiliation{Department of Physics and Astronomy, University of British Columbia, Vancouver, BC, V6T 1Z4 Canada}

\author{Piotr Sur\'owka}
\affiliation{Max-Planck-Institut f\"{u}r Physik komplexer Systeme, N\"{o}thnitzer Str. 38, 01187 Dresden, Germany}

\author{Vladimir Juri\v ci\' c}
\affiliation{Nordita, KTH Royal Institute of Technology and Stockholm University, Roslagstullsbacken 23,  10691 Stockholm,  Sweden}

\author{Bitan Roy}\email{bitan.roy@lehigh.edu}
\affiliation{Max-Planck-Institut f\"{u}r Physik komplexer Systeme, N\"{o}thnitzer Str. 38, 01187 Dresden, Germany}
\affiliation{Department of Physics, Lehigh University, Bethlehem, Pennsylvania, 18015, USA}

\date{\today}
\begin{abstract}
Electronic materials can sustain a variety of unusual, but symmetry protected touchings of valence and conduction bands, each of which is identified by a distinct topological invariant. Well-known examples include linearly dispersing pseudo-relativistic fermions in monolayer graphene, Weyl and nodal-loop semimetals, biquadratic (bicubic) band touching in bilayer (trilayer) graphene, as well as mixed dispersions in multi-Weyl systems. Here we show that depending on the underlying band curvature, the shear viscosity in the collisionless regime displays a unique power-law scaling with frequency at low temperatures, bearing the signatures of the band topology, which are distinct from the ones when the system resides at the brink of a topological phase transition into a band insulator. Therefore, besides the density of states (governing specific heat, compressibility) and dynamic conductivity, shear viscosity can be instrumental to pin nodal topology in electronic materials.
\end{abstract}

\maketitle

\begin{figure}[t!]
\includegraphics[width=0.95\linewidth, trim=0 2cm 0 2cm,clip]{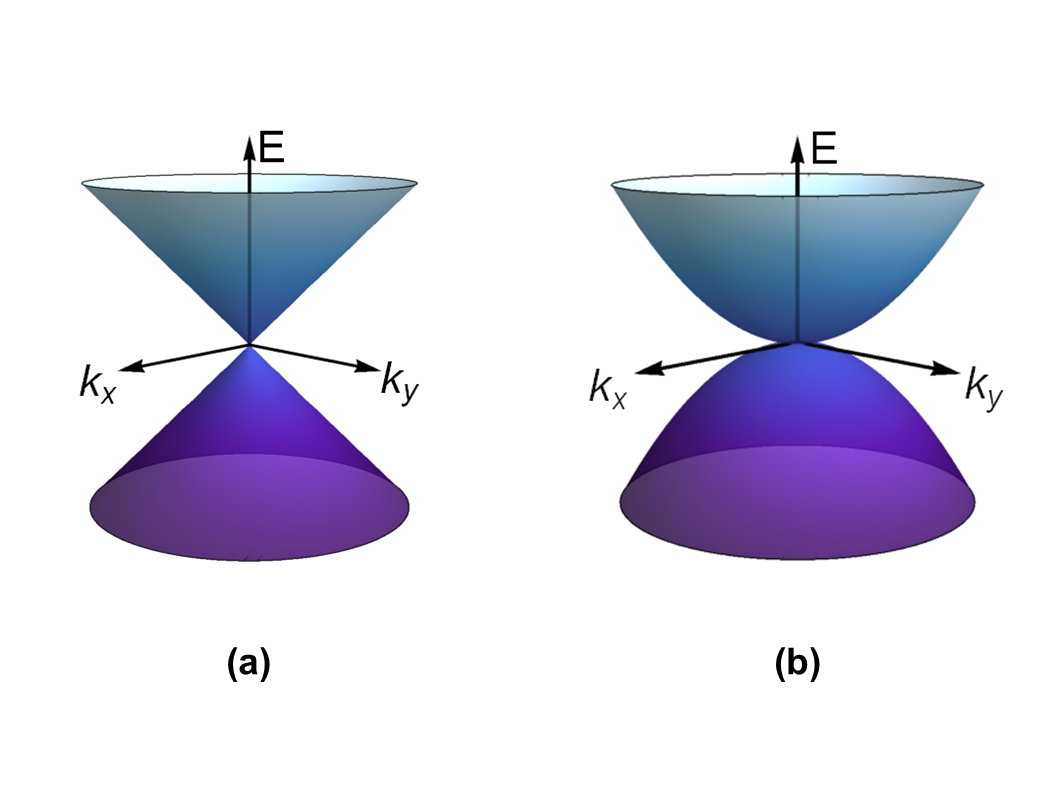}
\includegraphics[width=0.95\linewidth, trim=0 2cm 0 2cm,clip]{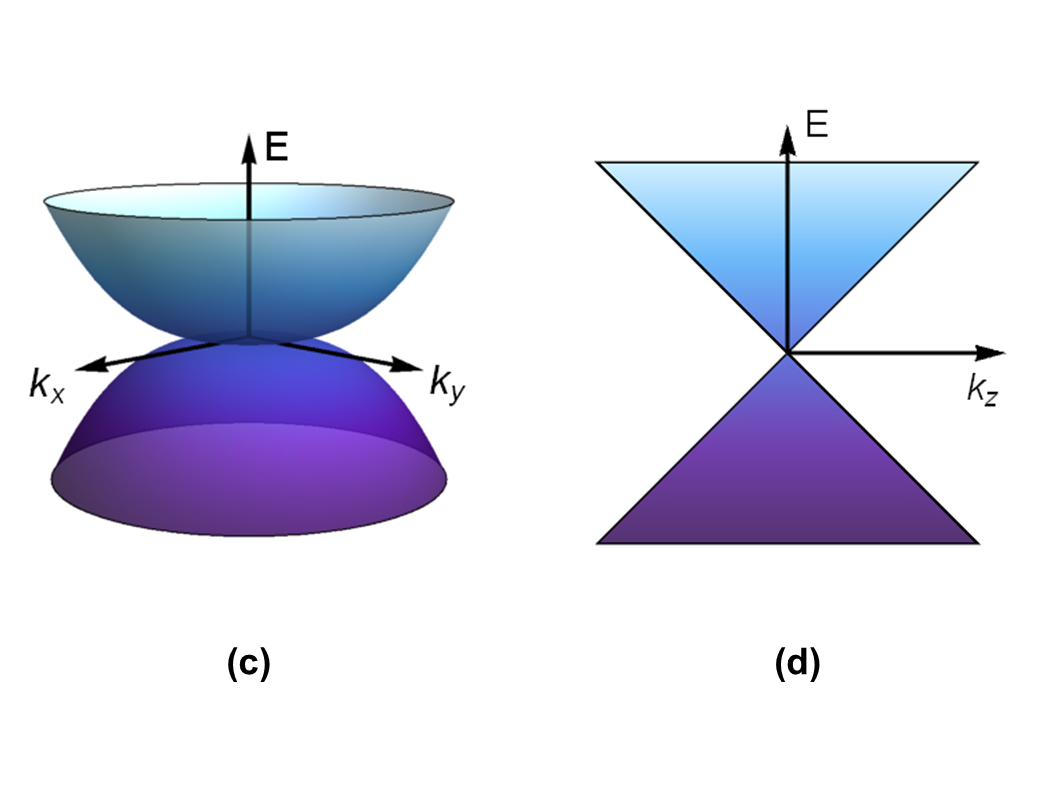}
\caption{Various topologically protected band touchings in two and three dimensions. In $d=2$, (a) linear, (b) biquadratic, and (c) bicubic touching of the conduction and valence bands can be realized in monolayer, Bernal bilayer and rhombohedral trilayer graphene, respectively. Energy dispersion in these systems respectively scales as $E \sim k^n_\perp$ with $n=1,2$ and $3$, where $k_\perp=\sqrt{k^2_x+k^2_y}$. In $d=3$, they correspond to multi-Weyl semimetals with monopole charge $n=1, 2$ and $3$, respectively. (d) For $k_\perp=0$, $E\sim k_z$ (always). In this construction, the Weyl nodes are separated along the $z$ direction.
}~\label{Fig:bandstructure}
\end{figure}

\emph{Introduction.~}A plethora of solid state compounds displays symmetry protected touchings of valence and conduction bands only at a few isolated points (nodes) in the reciprocal space~\cite{herring,luttinger, graphene-review, Weyl-review, balatsky, bradlyn-bernevig, schnyder-review}. They allow one to draw parallels with high-energy physics, when, in particular, emergent gapless quasiparticles are pseudo-relativistic in nature (due to the linear band dispersion), such as the ones found in atom-thick monolayer graphene and three-dimensional Weyl materials. The landscape of nodal Fermi liquids, however, encompasses territories that lie beyond the realm of standard Fermi liquids and relativistic systems, as the bands in the vicinity of the nodes can often acquire a finite \emph{curvature}. Known examples include the bi-quadratic (bi-cubic) touching in Bernal bilayer (rhombohedral trilayer) graphene~\cite{graphene-review}, and multi-Weyl semimetals displaying a distinct power-law dependence on different components of momenta~\cite{multiWeyl-1,multiWeyl-2,multiWeyl-3,multiWeyl-4,multiWeyl-5,multiWeyl-6,multiWeyl-7,multiWeyl-8,multiWeyl-9,multiWeyl-10, multiWeyl-11} (see Fig.~\ref{Fig:bandstructure}).

\begin{figure}[t!]
\includegraphics[width=0.95\linewidth]{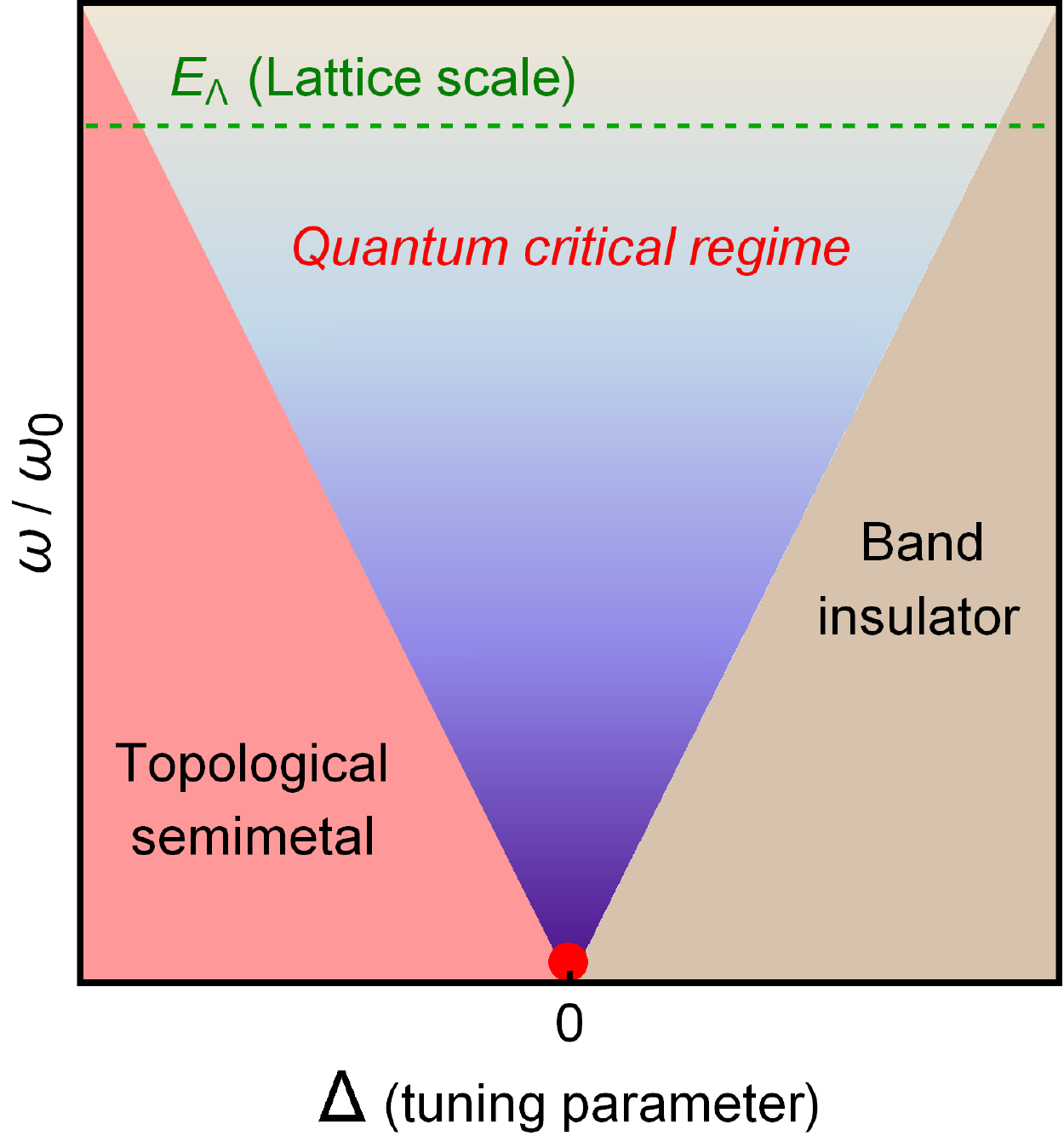}
\caption{A schematic phase diagram of two- and three-dimensional topological systems, residing at the verge of a quantum phase transition between a topological semimetal (pink region, see Fig.~\ref{Fig:bandstructure} for band dispersion) and a band insulator (dark beige region). The quantum critical point (red dot) separating these two symmetry indistinguishable phases is located at $\Delta=0$, and hosts an anisotropic semimetal. The proposed scaling of the shear viscosity for such a gapless phase can be observed over the quantum critical regime (purple shaded region) up to a nonuniversal lattice scale $E_\Lambda$ (green dashed line). Here, $\omega_0 \sim v^2_x/b$ and $(\alpha^2_n/b^n)^{1/(2-n)}$ in $d=2$ and $3$, respectively. In $d=2$ the topological semimetal always accommodates massless Dirac fermions (but, generically anisotropic), while in $d=3$ it represents a mWSM, possessing rotational symmetry in the $x-y$ plane and is characterized by an integer monopole charge $n$. The band insulator can be topological or trivial. At the TQCP (red dot) the in-plane dispersions are shown in panels (a)-(c) of Fig.~\ref{Fig:bandstructure}, while for $k_\perp=0$, $E\sim k_z^2$ in $d=3$. In $d=2$, the dispersion is linear along $k_x$ and quadratic along $k_y$ at the TQCP [see Eq.~(\ref{eq:2D-TQCP})].
}~\label{Fig:phasediagram}
\end{figure}

The nodes stand as \emph{topological defects} in the Brillouin zone and are characterized by distinct integer topological invariants, such as the vorticity in two dimensions or the monopole number in three dimensions. They in turn also fix the \emph{degeneracy} of robust metallic boundary modes: the bulk-boundary correspondence. Furthermore, by applying uniaxial strain or chemical pressure, for example, one can trigger \emph{collisions} between such topological defects, and place the system at the brink of band insulation~\cite{Montambaux-blochzener,opticallattice-1,opticallattice-2,blackphosphorus-1,roy-foster-PRX,roy-slager-juricic,roy-sur,saha-surowka,Muralidharan,hasan-topotransition,landsteiner-Topotrans,kruger}. The system at such topological quantum critical points (TQCPs) describes an infrared unstable phase of matter, as shown in Fig.~\ref{Fig:phasediagram}. Hence, the kingdom of nodal Fermi liquids is constituted by a diverse collection of unconventional gapless fermionic systems, and their characterization in terms of a bulk response (the optical shear viscosity) is the central theme of this work.

Even though the thermodynamic quantities, such as specific heat ($C_v$), compressibility ($\kappa$), and the dynamic conductivity (${\hat \sigma}$) provide valuable insights into the band curvature (and henceforth the nodal topology), they also suffer some limitations. For example, owing to the $|E|$-linear density of states, $C_v \sim T^2$ and $\kappa \sim T$ in monolayer graphene, and three-dimensional double-Weyl and nodal-loop semimetals, at sufficiently low temperatures ($T$). On the other hand, in any two-dimensional \emph{isotropic} system the diagonal conductivity $\sigma\sim e^2/h$, due to gauge invariance. Therefore, unambiguous characterization of nodal topology demands new bulk probes, which we attempt to meet by focusing on the \emph{shear viscosity} in the collisionless or high-frequency regime ($\hbar \Omega \gg k_B T$). We show that the shear viscosity $\eta_{\alpha \beta \gamma \delta} (\Omega)$ displays a richer power-law scaling with frequency ($\Omega$), as it depends on \emph{four} spatial indices. To this end, we employ the Kubo formalism and set $T=0$ from the outset. Throughout this Rapid Communication we set $\hbar=1$ and our main results are the following.

Based on standard linear response theory, we formulate a unifying approach for addressing elastic responses in a broad class of anisotropic nodal semimetals. Concretely, our methodology encompasses two- and three-dimensional (2D and 3D) semimetallic systems featuring: $(i)$ restricted in-plane rotational symmetry, and $(ii)$ higher topological winding number or pseudospin angular momentum. This development is based on a general definition for the stress tensor [Eqs.~(\ref{eq:T-general}) and~(\ref{eq:Torbital})], together with a careful identification of the generators of rotations in the systems with restricted rotational symmetry [Eq.~\eqref{eq:Tpseudospin}]. Within this framework, we compute the high-frequency (optical) shear viscosity in different classes of topological semimetals characterized by vorticity (monopole charge) in two (three) dimensions, as well as in infrared unstable semimetals, occupying topological quantum critical regimes. In two spatial dimensions all components of the viscosity tensor $\eta_{ijkl} \sim \Omega^{2/n}$ ($i,j,k,l=x,y$), where $n$ counts the momentum-space vorticity of the nodes (topological invariant). Respectively, $n=1,2$ and $3$ correspond to linear, biquadratic and bicubic band touching of chiral fermions [Fig.~\ref{Fig:bandstructure}], relevant for monolayer, Bernal bilayer and rhombohedral trilayer graphene. By contrast, in three-dimensional multi-Weyl semimetals (mWSMs) 
\begin{equation}
\eta_{_\perp}, \eta_{j j z z}, \eta_{zzzz} \sim \Omega^{\frac{2}{n}+1},\,\,\eta_{jzjz} \sim \Omega^3,\,\, 
\eta_{zjzj} \sim \Omega^{\frac{4}{n}-1}, \nonumber 
\end{equation}
with $\eta_{_\perp}$ as the in-plane ($x-y$) components of the viscosity tensor. Therefore, the optical shear viscosity can directly probe the topological charge ($n$) of the bulk Weyl node. Furthermore, this observable can also be instrumental to identify TQCPs, separating a bulk semimetal from a band insulator, see Fig.~\ref{Fig:phasediagram}. Specifically, in two dimensions
\begin{equation} 
\eta_{j j j j},\eta_{j j k k},
\eta_{j k k j} \sim \Omega^{\frac{3}{2}},\,\, \eta_{x y x y}\sim \Omega^{\frac{1}{2}},\,\, \eta_{y x y x}\sim\Omega^{\frac{5}{2}}, \nonumber
\end{equation} 
which reflects the absence of rotational symmetry at this TQCP. On the other hand, in three dimensions
\allowdisplaybreaks[4]
\begin{equation}
\eta_{_\perp}, \eta_{z z z z}, \eta_{z z j j}, \eta_{z j j z}\sim \Omega^{\frac{2}{n}+\frac{1}{2}},\,
\eta_{j z j z}\sim \Omega^{\frac{3}{2}},\, \eta_{z j z j}\sim \Omega^{\frac{4}{n}-\frac{1}{2}}, \nonumber
\end{equation}
due to the invariance of the system under in-plane rotations about the $z$ axis. Electronic viscosity has so far been measured in the hydrodynamic regime~\cite{viscosity-exp-1, viscosity-exp-2,viscosity-exp-3}, as well as at low temperature in graphene~\cite{grahene-viscosity-lowT}. The present Rapid Communication should motivate similar experiments in topological electronic fluids, but in the optical or collisionless regime, when the rate of deformation of the system is much larger than temperature, i.e. $\hbar \Omega \gg k_B T$.

\emph{General setup.~}Here we consider  the dynamical (optical) shear viscosity in the collisionless regime at finite frequency ($\Omega$) and zero temperature, which is the response of an elastic medium to a time-dependent (dynamic) volume-preserving (shear) strain perturbation in terms of the stress tensor. The stress tensor is defined from the Hamiltonian operator $H({\bf k})$ according to~\cite{bradlyn-read,patel-sachdev,Link-1,Landsteiner-viscosity,Link-2,Rao-Bradlyn}
\begin{equation}~\label{eq:T-general}
T_{\alpha\beta}({\bf k})=-i[H({\bf k}),{\mathcal J}_{\alpha\beta}],
\end{equation}
where ${\mathcal J}_{\alpha\beta}={\mathcal L}_{\alpha\beta}+{\mathcal S}_{\alpha\beta}$ is the total angular momentum, ${\mathcal L}_{\alpha\beta}=-x_\alpha k_\beta$ is the orbital angular momentum, while ${\mathcal S}_{\alpha\beta}$ generates pseudo-spin rotations~\cite{Link-1}, with the form that depends on the system, as specified below. Here, ${\bf x}$ and ${\bf k}$ are the position and the momentum operators, respectively. It can be readily shown that the orbital part of the stress tensor has the form 
\begin{equation}~\label{eq:Torbital}
T^{\rm (o)}_{\alpha\beta}({\bf k})=k_\beta \; \frac{\partial H ({\bf k})}{\partial k_\alpha}.
\end{equation}
It is important to emphasize that here we mostly consider systems that feature low energy fermionic quasiparticles with a reduced rotational symmetry [a subgroup of the full SU(2) group]. Thus, the corresponding spinor field transforms nontrivially in the pseudospin space only under this reduced rotational symmetry. 

The Kubo formula for the viscosity tensor is then given in terms of the stress tensor correlation function 
\begin{equation}\label{eq:Kubo-eta}
\eta_{\alpha\beta\gamma\delta}(\Omega)=\frac{{\rm Im} \: C_{\alpha\beta\gamma\delta}(i\Omega\to\Omega+i\delta)}{\Omega},
\end{equation}
where in $d$ spatial dimensions
\begin{align}\label{eq:TT-correlator}
C_{\alpha\beta\gamma\delta}(i\Omega)&=-\int_{-\infty}^{\infty}\frac{d\omega}{2\pi}\int\frac{d^d{\bf k}}{(2\pi)^d}
{\rm Tr}[G_F(i\omega,{\bf k})T_{\alpha\beta}({\bf k})\nonumber\\
&\times G_F(i\omega+i\Omega,{\bf k})T_{\gamma\delta}({\bf k})].
\end{align}
Notice that to obtain the viscosity tensor, the  stress tensor correlator $C_{\alpha\beta\gamma\delta}(i\Omega)$ at Matsubara frequency $i\Omega$ has to be analytically continued to real frequency $i\Omega\to\Omega+i\delta$, with $\delta\to0$. The fermion propagator in terms of the corresponding Hamiltonian $H({\bf k})$ reads as
\begin{equation}\label{eq:GreensF}
G_F(i\omega,{\bf k})=\frac{i\omega+H({\bf k})}{\omega^2+H^2({\bf k})}. 
\end{equation}
The scaling dimension of the viscosity tensor in units of momentum (inverse length) can be obtained from the scaling dimension of the stress tensor, given by Eq.~\eqref{eq:Torbital}, from which we find that ${\rm dim}[T_{\alpha\beta}]={\rm dim}[\Omega]$. Since ${\rm dim}[G_F]=-{\rm dim}[\Omega]$ [see Eq.~\eqref{eq:GreensF}], the term under the trace in Eq.~\eqref{eq:TT-correlator} is dimensionless, yielding ${\rm dim}[C_{\alpha\beta\gamma\delta}]=d+{\rm dim}[\Omega]$. From \eqref{eq:Kubo-eta}, we then obtain ${\rm dim}[\eta_{\alpha\beta\gamma\delta}]=d$. Furthermore, we point out that for an isotropic and rotationally symmetric medium in $d$ spatial dimensions the viscosity tensor is characterized by only \emph{one} independent component $\eta(\Omega)$, given by~\cite{Landau} 
\begin{equation}~\label{eq:viscosity-RS}
\eta_{\alpha\beta\gamma\delta}(\Omega)=\eta(\Omega) \; {\cal P}_{\alpha\beta\gamma\delta},
\end{equation}
where ${\cal P}_{\alpha\beta\gamma\delta}=\delta_{\alpha\gamma}\delta_{\beta\delta}+\delta_{\alpha\delta}\delta_{\beta\gamma}-(2/d)\delta_{\alpha\beta}\delta_{\gamma\delta}$, and $\delta_{\alpha\beta}$ is the Kronecker $\delta$ symbol. In general, however, the number of independent components of this tensor depends on the dimensionality and the symmetry of the system. It can also be explicitly seen for different classes of nodal semimetals, discussed below.

\emph{Topological~semimetals.}~We first consider two-dimensional topological semimetals, described by the minimal model in the vicinity of the band touching points
\begin{equation}~\label{eq:master-Ham-mWSM}
H_{n}^{\rm 2D}({\bf k})=\alpha_n k_\perp^n \left[ \cos (n\phi) \sigma_x + \sin (n\phi) \sigma_y \right],
\end{equation}
at low-energies, where $\phi=\tan^{-1}(k_y/k_x)$, $0\leq \phi<2\pi$, $k_\perp^2=k_x^2+k_y^2$, and $n$ counts the momentum-space vorticity. The Pauli matrices ${\bm\sigma}$ act in the pseudospin space~\cite{graphene-review}. Notice that ${\rm dim}[\alpha_n]={\rm dim}[\Omega]-n$, and $\alpha_1$ bears the dimension of Fermi velocity (dimensionless), while $\alpha_2$ has the dimension of the inverse mass. This system possesses U(1) rotational symmetry, with the orbital part of the stress tensor given by Eq.~\eqref{eq:Torbital}. The pseudospin part is obtained from Eq.~\eqref{eq:T-general} by noting that the pseudospin rotations are generated by 
\begin{equation}~\label{eq:Tpseudospin}
{\mathcal S}_{i j}=-\frac{n}{4} \epsilon_{i j z} \sigma_z, 
\end{equation}
with $\epsilon_{i j k}$ being the fully antisymmetric Levi-Civita symbol. After computing the components of the stress tensor and using the Kubo formula, the viscosity tensor has the general form for an isotropic and rotationally symmetric system, displayed in Eq.~\eqref{eq:viscosity-RS}, with $d=2$ and
\allowdisplaybreaks[4]
\begin{equation}~\label{eq:viscosity-2D-SM}
\eta (\Omega) \equiv \eta(\Omega,n)=\frac{n}{64}\left(\frac{\Omega}{2\alpha_n}\right)^{\frac{2}{n}},
\end{equation}
which explicitly depends on the vorticity ($n$) of the band touching point. See Supplemental Materials for details~\cite{supplemental}. Therefore, with increasing vorticity two-dimensional topological semimetals become more \emph{viscous}. 

In three dimensions, low-energy excitations in a mWSM near one of the Weyl points, possessing monopole charge $n$, with $n>0$, are described by the Hamiltonian from Eq.~\eqref{eq:master-Ham-mWSM}, together with an extra term describing linearly dispersing quasiparticles along the separation of Weyl nodes, the $z$-axis~\cite{multiWeyl-1, multiWeyl-2, multiWeyl-3, multiWeyl-4, multiWeyl-6, multiWeyl-9, multiWeyl-10}
\begin{equation}~\label{eq:Ham-mWSM}
H_n^{\rm 3D}({\bf k})=H_n^{\rm 2D}(k_x,k_y)+v_z k_z \sigma_z.
\end{equation}
The system is rotationally invariant only in the $x-y$ plane for any $n>1$, while the full rotational symmetry is restored for the simple WSM ($n=1$). The orbital part of the stress tensor is given by Eq.~\eqref{eq:Torbital}, while the pseudospin part is obtained from the generator of the pseudospin rotations about the $z$-axis, given by Eq.\eqref{eq:Tpseudospin}.~The resulting in-plane components of the viscosity tensor assume the form in Eq.~\eqref{eq:viscosity-RS}, with a monopole charge-dependent effective dimensionality associated with the shear viscosity~$d\equiv d_\star(n)=(4n+2)/(n+1)$, and
\begin{equation}\label{eq:viscosity-xy-mWSM}
\eta=\frac{n^3 \Gamma\left(4+\frac{1}{n}\right)}{2^{8+\frac{2}{n}} (n+1)(3n+1)\sqrt{\pi}\Gamma\left(\frac{5}{2}+\frac{1}{n}\right)}
\left(\frac{\Omega}{v_z}\right)\left(\frac{\Omega}{\alpha_n}\right)^{\frac{2}{n}},
\end{equation}
where $\eta\equiv\eta(\Omega, n)$ and $\Gamma(x)$ is the gamma function. The remaining components involving the $z$-direction are 
\begin{align}
\eta_{zzzz}&=-\frac{2}{n}\eta_{jjzz}=\frac{4}{n(2n+1)}\eta(\Omega,n),\\
\eta_{jzjz}&=\frac{n}{240\pi}\left(\frac{\Omega}{v_z}\right)^3+\frac{\Omega^3(v_z-2\alpha_1)}{384\pi\alpha_1^2 v_z^2}\delta_{n,1},
\label{eq:etajzjz}\\
\eta_{jzzj}&=\frac{\Gamma\left(1+\frac{1}{n}\right)}{2^{7+\frac{2}{n}}\sqrt{\pi}\Gamma\left(\frac{5}{2}+\frac{1}{n}\right)}\frac{\Omega}{v_z}\left(\frac{\Omega}{\alpha_n}\right)^{2/n}\nonumber\\
&+\frac{1}{384\pi}\left(\frac{v_z}{\alpha_1}+\frac{\alpha_1}{v_z}-1\right)\frac{\Omega}{v_z}\left(\frac{\Omega}{\alpha_1}\right)^2\delta_{n,1},~\label{eq:etajzzj}\\
\eta_{zjzj}&=F(n)\left(\frac{\Omega}{\alpha_n}\right)^{\frac{4}{n}}\left(\frac{\Omega}{v_z}\right)^{-1}+
\frac{\Omega^3(\alpha_1-2v_z)}{384\pi\alpha_1^3 v_z}\delta_{n,1}~\label{eq:etazjzj},
\end{align}
where $F(n)=\Gamma(2/n)/[2^{3+4/n}n^2\sqrt{\pi}\Gamma(3/2+2/n)]$ and $\eta\equiv\eta(\Omega,n)$, see Eq.~\eqref{eq:viscosity-xy-mWSM}.~The simple WSM deserves special attention, since it features full SU(2) rotational symmetry, with ${\cal S}_{\alpha\beta}=(i/8)[\sigma_\alpha,\sigma_\beta]$ as the generators of the pseudospin rotations, which yield extra contributions to the viscosity tensor, see Eqs.~\eqref{eq:etajzjz}-\eqref{eq:etazjzj}. Furthermore, when in addition the in-plane and out-of-plane velocities are equal, $\alpha_1=v_z\equiv v$, the system is also fully isotropic and the viscosity acquires the form in Eq.~\eqref{eq:viscosity-RS}, with $d=d_\star(n=1)=3$ and $\eta(\Omega,n=1)=(\Omega/v)^3/(640 \pi)$.

\emph{Nodal semimetals at TQCPs.~} We now examine the shear viscosity when a bulk nodal semimetal is at the brink of the topological quantum phase transition (tuned by $\Delta$) into a band insulator, see Fig.~\ref{Fig:phasediagram}. We first consider a two-dimensional Hamiltonian 
\begin{equation}~\label{eq:2D-TQCP}
H_{\rm TQCP}^{\rm 2D}({\bf k})=v_x k_x \sigma_x +(b k_y^2+\Delta)\sigma_y,
\end{equation}
describing a symmetry indistinguishable Dirac semimetal and a band insulator, respectively for $\Delta<0$ and $\Delta>0$. They are separated by a TQCP, hosting an anisotropic semimetal and located at $\Delta=0$~\cite{Montambaux-blochzener,opticallattice-1, opticallattice-2,blackphosphorus-1, roy-foster-PRX, saha-surowka, Muralidharan,roy-sur, kruger}. Here, the velocity $v_x>0$ and the band curvature in the $y-$direction $b=(1/4){\rm Tr}[\sigma_y\partial^2H/\partial k_y^2]>0.$ Notice that  irrespective of the vorticity, there is only one type of topological quantum critical theory in two spatial dimensions, since for any $n>1$ the nodal semimetal-band insulator transition occurs in two steps. First, a vorticity-$n$ nodal point splits into $n$ copies of simple linearly dispersing (but anisotropic) points with $n=1$, which then collide with each other, yielding the band insulator. The latter transition is described by the Hamiltonian in Eq.~\eqref{eq:2D-TQCP}. Notice that rotational symmetry is completely broken in this process, reflecting the absence of such a symmetry in the corresponding Hamiltonian operator. Therefore, the stress tensor is solely given by the orbital part $T^{\rm (o)}_{\alpha\beta}({\bf k})$, with four independent components~\cite{Link-2}. In turn, we obtain six independent components of the viscosity tensor
\begin{align}~\label{eq:viscosity2D-TQCP}
\eta_{xxxx}&=-\frac{\eta_{xxyy}}{2}=-\frac{\eta_{xyyx}}{2}=\frac{\eta_{yyyy}}{4}=
\frac{E_K\left(\frac{1}{2}\right)}{336\pi} \sqrt{\frac{\Omega^3}{v_x^2 b}}, \nonumber \\
\eta_{xyxy}&=\frac{9v^4}{4\pi b^2\Omega^2}\eta_{yxyx}=\frac{3\left[\Gamma\left(-\frac{1}{4}\right)\right]^2}{1280\pi^{3/2}}
\left(\frac{v_x\sqrt{\Omega}}{b^{3/2}}\right),
\end{align}
indeed discerning strong anisotropy due to the absence of rotational symmetry at the TQCP in two dimensions. Here $E_K(x)$ is the elliptic function of the first kind. This outcome should be contrasted to the isotropic and rotationally symmetric semimetallic phases, with a single viscosity coefficient, shown in Eq.~\eqref{eq:viscosity-2D-SM}. Finally, we note that with decreasing band curvature $b$, all components of the viscosity tensor increase at the TQCP.

Next we focus on the Hamiltonian~\cite{multiWeyl-3, multiWeyl-6, roy-slager-juricic, landsteiner-Topotrans} 
\begin{equation}~\label{eq:3D-TQCP}
H_{\rm TQCP}^{\rm 3D}({\bf k})=H_n^{\rm 2D}(k_x,k_y)+(b k_z^2+\Delta)\sigma_z,
\end{equation}
describing a TQCP ($\Delta=0$),~separating a three-dimensional mWSM ($\Delta<0$) and a symmetry indistinguishable band insulator ($\Delta>0$), see Fig.~\ref{Fig:phasediagram}. As the Weyl points are separated along the $z$-direction, the quadratic dispersion at the TQCP appears in this direction. Topological quantum critical theories in three dimensions are therefore monopole charge-dependent, reflecting the in-plane rotational symmetry, in contrast to the situation in two dimensions, where such a symmetry is absent, yielding a unique theory describing such a transition.

Due to the presence of in-plane pseudospin rotational symmetry for any $n$, with the same generator as for the Weyl semimetallic phase, the in-plane components of the viscosity tensor are given by Eq.~\eqref{eq:viscosity-RS}, with the effective dimensionality associated with shear viscosity at the TQCP $d\equiv d_{\star,{\rm TQCP}}(n)=(3n+2)/(n+1)$, and
\begin{equation}~\label{eq:viscosity-xy-TQCP}
\eta_{_{\rm TQCP}}=\frac{(3n+2)\Gamma\left(\frac{5}{4}\right) \Gamma\left(1+\frac{1}{n}\right)}{2^{8+2/n}\sqrt{2}\;\pi
\Gamma\left(\frac{9}{4}+\frac{1}{n}\right)}
\sqrt{\frac{\Omega}{b}}\left(\frac{\Omega}{\alpha_n}\right)^{\frac{2}{n}}.
\end{equation}
Notice that the effective dimensionality at the TQCP is different than that for the bulk phase due to the difference in the quasiparticle dispersion in the $z$-direction in these two cases. The remaining components of the viscosity tensor involving the $z$-direction are obtained only from the orbital part of the stress tensor, and given by
\begin{align}~\label{eq:viscosity-other-TQCP}
\eta_{zzzz}&=-\frac{4}{n}\eta_{zzjj}=\frac{16 \; \eta_{_{\rm TQCP}}}{n(3n+2)}, 
\eta_{jzjz}=\frac{5n}{672\sqrt{2}\pi}\left(\frac{\Omega}{b}\right)^{\frac{3}{2}},
\nonumber \\
\eta_{zjzj}&=\frac{\Gamma\left(\frac{3}{4}\right)\Gamma\left(\frac{2}{n}\right)}{2^{3+4/n}\sqrt{2}\pi 
\Gamma\left(\frac{7}{4}+\frac{2}{n}\right)n^2}\left(\frac{\Omega}{b}\right)^{-\frac{1}{2}}
\left(\frac{\Omega}{\alpha_n}\right)^{\frac{4}{n}}, \nonumber \\
\eta_{jzzj}&=-\frac{\Gamma\left(\frac{5}{4}\right)\Gamma\left(1+\frac{1}{n}\right)}{64\pi 2^{2/n}\Gamma\left(\frac{9}{4}+\frac{1}{n}\right)}
\left(\frac{\Omega}{b}\right)^{\frac{1}{2}}\left(\frac{\Omega}{\alpha_n}\right)^{\frac{2}{n}}.
\end{align}
Notice that all components of the viscosity tensor explicitly depend on the monopole charge ($n$), but this dependence differs in the cases of the bulk semimetal and the infrared unstable gapless state at the TQCP. Analogously to the $d=2$ case, in $d=3$ all components of the viscosity tensor also increase as the band curvature along the $z$ direction ($b$) decreases.

\emph{Discussion}. We develop a general method for computing the elastic responses, and in particular, the shear viscosity in the collisionless or high-frequency regime, in a wide class of nodal semimetals that display invariance under a restricted rotational symmetry and/or anisotropic quasiparticle spectra [see Eqs.~(\ref{eq:T-general}),~(\ref{eq:Torbital}) and~(\ref{eq:Tpseudospin})]. We show that in the former class of systems, the in-plane components of the viscosity assume the form of those in isotropic systems [see Eq.~\eqref{eq:viscosity-RS}], with the physical dimension ($d$) being replaced by an appropriate \emph{effective} dimension ($d_\star$). Specifically in three-dimensional Weyl materials, $d_\star$ depends on the monopole number ($n$), see $d_\star(n)$ and $d_{\star,{\rm TQCP}}(n)$. This approach is also applicable to the nodal semimetals with a quasiparticle spectrum that depends on an arbitrary power of momentum in both two and three spatial dimensions. We show that the scaling of the optical shear viscosity with frequency can be instrumental for probing the vorticity [Eq.~(\ref{eq:viscosity-2D-SM})] and the monopole charge [Eqs.~(\ref{eq:viscosity-xy-mWSM})-(\ref{eq:etazjzj})] of bulk nodal semimetals in two and three dimensions, respectively. Furthermore, the scaling of this observable and particularly its anisotropy can be used to detect TQCPs separating a nodal semimetal from a band insulator [Eq.~(\ref{eq:viscosity2D-TQCP}),~(\ref{eq:viscosity-xy-TQCP}), and~(\ref{eq:viscosity-other-TQCP})]. We hope that predicted scaling of the shear viscosity will motivate future experiments in the optical or collisionless regime~\cite{viscosity-exp-1, viscosity-exp-2,viscosity-exp-3,grahene-viscosity-lowT}.

Our formalism can be generalized to other classes of systems with different types of band touching points, such as bi-quadratic Luttinger fermions in three dimensions~\cite{luttinger, link-herbut}, birefringent and multirefringent fermions which realize higher spin fermions in two and three dimensions~\cite{roy-birefringent-PRL}, as well as nodal loop semimetals~\cite{nodal-loop-broy}. This formalism can be extended to incorporate the effects of electronic interactions and disorder on the shear viscosity in topological and quantum critical~\cite{maciejko} materials. Shear viscosity could also be used for probing gapless superconducting states, in general displaying anisotropic low energy dispersion of the Bogoliubov-de Gennes quasiparticles~\cite{volovik, sigrist}, which we leave for a future investigation.

\emph{Acknowledgment.}~M.M. was supported by NSERC and FRQNT (Grant No.~273327). P.S. was supported by the Deutsche Forschungsgemeinschaft through the Leibniz Program and the cluster of excellence ct.qmat (EXC 2147, project-ID 39085490). V.J. acknowledges the support from the Swedish Research Council (VR 2019-04735). B.R. was supported by the Startup Grant from Lehigh University.


\begin{thebibliography}{}

\bibitem{herring} C. Herring, Phys. Rev. {\bf 52}, 365 (1937).

\bibitem{luttinger} J. M. Luttinger, Phys. Rev. {\bf 102}, 1030 (1956).

\bibitem{graphene-review} A. H. Castro Neto, F. Guinea, N. M. R. Peres, K. S. Novoselov, and A. K. Geim, Rev. Mod. Phys. {\bf 81}, 109 (2009).

\bibitem{Weyl-review} N. P. Armitage, E. J. Mele, and A. Vishwanath, Rev. Mod. Phys. {\bf 90}, 15001 (2018).

\bibitem{balatsky} T. O. Wehling, A. M. Black-Schaffer, and A. V. Balatsky, Adv. Phys. {\bf 63}, 1 (2014).

\bibitem{bradlyn-bernevig} B. Bradlyn, J. Cano, Z. Wang, M. G. Vergniory, C. Felser, R. J. Cava, and B. A. Bernevig, Science {\bf 353}, aaf5037 (2016).

\bibitem{schnyder-review} C.-K. Chiu, J. C. Y. Teo, A. P. Schnyder, and S. Ryu, Rev. Mod. Phys. {\bf 88}, 035005 (2016).

\bibitem{multiWeyl-1} G. Xu, H. Weng, Z. Wang, X. Dai, and Z. Fang, Phys. Rev. Lett. {\bf 107}, 186806 (2011).

\bibitem{multiWeyl-2} C. Fang, M. J. Gilbert, X. Dai, and B. A. Bernevig, Phys. Rev. Lett. {\bf 108}, 266802 (2012).

\bibitem{multiWeyl-3} B.-J. Yang and N. Nagaosa, Nat. Commun. {\bf 5}, 4898 (2014).

\bibitem{multiWeyl-4} B. Roy and J. D. Sau, Phys. Rev. B {\bf 92}, 125141 (2015).

\bibitem{multiWeyl-5} S.-M. Huang, S.-Y. Xu, I. Belopolski, C.-C. Lee, G. Chang, T.-R. Chang, B. Wang, N. Alidoust, G. Bian, M. Neupane, D. Sanchez, H. Zheng, H.-T. Jeng, A. Bansil, T. Neupert, H. Lin, and M. Z. Hasan, Proc. Natl. Acad. Sci. USA {\bf 113}, 1180 (2016).

\bibitem{multiWeyl-6} B. Roy, P. Goswami, and V. Juri\v ci\' c, Phys. Rev. B {\bf 95}, 201102(R) (2017).

\bibitem{multiWeyl-7} Q. Liu and A. Zunger, Phys. Rev. X {\bf 7}, 021019 (2017).

\bibitem{multiWeyl-8} S-K. Jian and H. Yao, Phys. Rev. B {\bf 96}, 155112 (2017).

\bibitem{multiWeyl-9} R. M. A. Dantas, F. Pe\~{n}a-Benitez, B. Roy and P. Sur\'owka, Phys. Rev. Research {\bf 2}, 013007 (2020). 

\bibitem{multiWeyl-10} R. Soto-Garrido, E. Mu\~{n}oz, and V. Juri\v ci\' c, Phys. Rev. Research {\bf 2}, 012043 (2020).

\bibitem{multiWeyl-11} S. Nandy, S. Manna, D. C\u{a}lug\u{a}ru, and B. Roy, Phys. Rev. B {\bf 100}, 235201 (2019). 

\bibitem{Montambaux-blochzener} L-K. Lim, J-N. Fuchs, and G. Montambaux, Phys. Rev. Lett. {\bf 108}, 175303 (2012).

\bibitem{opticallattice-1} L. Tarruell, D. Greif, T. Uehlinger, G. Jotzu, and T. Esslinger, Nature (London) {\bf 483}, 302 (2012).

\bibitem{opticallattice-2} M. Tarnowski, M. Nuske, N. Fl\"aschner, B. Rem, D. Vogel, L. Freystatzky, K. Sengstock, L. Mathey, and C. Weitenberg, Phys. Rev. Lett. {\bf 118}, 240403 (2017).

\bibitem{blackphosphorus-1} J. Kim, S. S. Baik, S. H. Ryu, Y. Sohn, S. Park, B-G. Park, J. Denlinger, Y. Yi, H. J. Choi, and K. S. Kim, Science {\bf 349}, 723 (2015).

\bibitem{roy-foster-PRX} B. Roy and M. S. Foster, Phys. Rev. X {\bf 8}, 011049 (2018).

\bibitem{saha-surowka} F. Pe\~{n}a-Benitez, K. Saha, and P. Sur\'owka, Phys. Rev. B {\bf 99}, 045141 (2019). 

\bibitem{Muralidharan} A. Mawrie and B. Muralidharan, Phys. Rev. B {\bf 99}, 075415 (2019).

\bibitem{roy-sur} S. Sur and B. Roy, Phys. Rev. Lett. {\bf 123}, 207601 (2019).

\bibitem{kruger} M. D. Uryszek, E. Christou, A. Jaefari, F. Kr\"uger, and B. Uchoa, Phys. Rev. B {\bf 100}, 155101 (2019).

\bibitem{roy-slager-juricic} B. Roy, R-J. Slager, and V. Juri\v ci\' c, Phys. Rev. X {\bf 8}, 031076 (2018).

\bibitem{hasan-topotransition} S-Y. Xu, M. Neupane, I. Belopolski, C. Liu, N. Alidoust, G. Bian, S. Jia, G. Landolt, B. Slomski, J. H. Dil, P. P. Shibayev, S. Basak, T-R. Chang, H-T. Jeng, R. J. Cava, H. Lin, A. Bansil, M. Z. Hasan, Nat. Commun. {\bf 6}, 6870 (2015).

\bibitem{landsteiner-Topotrans} K. Landsteiner, Y. Liu, and Y-W. Sun, Sci. China Phys. Mech. Astron. {\bf 63}, 250001 (2020).


\bibitem{viscosity-exp-1} D. A. Bandurin, I. Torre, R. K. Kumar, M. B. Shalom, A. Tomadin, A. Principi, G. H. Auton, E. Khestanova, K. S. Novoselov, I. V. Grigorieva, L. A. Ponomarenko, A. K. Geim, M. Polini, Science {\bf 351}, 1055 (2016).

\bibitem{viscosity-exp-2} J. Crossno, J. K. Shi, K. Wang, X. Liu, A. Harzheim, A. Lucas, S. Sachdev, P. Kim, T. Taniguchi, K. Watanabe, T. A. Ohki, K.-C. Fong, Science {\bf 351}, 1058 (2016). 

\bibitem{viscosity-exp-3} P. J. W. Moll, P. Kushwaha, N. Nandi, B. Schmidt, A. P. Mackenzie, Science {\bf 351}, 1061 (2016).

\bibitem{grahene-viscosity-lowT} R. Krishna Kumar, D. A. Bandurin, F. M. D. Pellegrino, Y. Cao, A. Principi, H. Guo, G. H. Auton, M. Ben Shalom, L. A. Ponomarenko, G. Falkovich, K. Watanabe, T. Taniguchi, I. V. Grigorieva, L. S. Levitov, M. Polini, and A. K. Geim, Nat. Phys. {\bf 13}, 1182 (2017).

\bibitem{bradlyn-read} B. Bradlyn, M. Goldstein, and N. Read, Phys. Rev. B {\bf 86}, 245309 (2012).

\bibitem{patel-sachdev} A. Eberlein, A. Patel, and S. Sachdev, Phys. Rev. B {\bf 95}, 075127 (2017).

\bibitem{Link-1} J. M. Link, D. E. Sheehy, B. N. Narozhny, and J. Schmalian, Phys. Rev. B {\bf 98}, 195103 (2018).

\bibitem{Landsteiner-viscosity} C. Copetti and K. Landsteiner, Phys. Rev. B {\bf 99}, 195146 (2019).

\bibitem{Link-2} J. M. Link, B. N. Narozhny, E. I. Kiselev, and J. Schmalian, Phys. Rev. Lett. {\bf 120}, 196801 (2018).

\bibitem{Rao-Bradlyn} P. Rao and B. Bradlyn, Phys. Rev. X {\bf 10}, 021005 (2020). 

\bibitem{Landau} L. D. Landau and E. M. Lifshitz, \emph{Fluid Mechanics} (Butterworth-Heinemann, Oxford, UK, 2000). 

\bibitem{supplemental} See Supplemental Materials at XXX-XXXX for technical details involved in the computation of the viscosity tensor.

\bibitem{link-herbut} J. M. Link and I. F. Herbut, Phys. Rev. B {\bf 101}, 125128 (2020).

\bibitem{roy-birefringent-PRL} B. Roy, M. P. Kennett, K. Yang, and Juri\v ci\' c, Phys. Rev. Lett. {\bf 121}, 157602 (2018); B. Roy and V. Juri\v ci\' c, Phys. Rev. Research {\bf 2}, 012047 (2020). 

\bibitem{nodal-loop-broy} B. Roy, Phys. Rev. B {\bf 96}, 041113(R) (2017).

\bibitem{maciejko} W. Witczak-Krempa and J. Maciejko, Phys. Rev. Lett. {\bf 116}, 100402 (2016).

\bibitem{volovik} G. E. Volovik, \emph{The Universe in a Helium Droplet} (Oxford University Press, Oxford, UK, 2003).

\bibitem{sigrist} M. Sigrist and K. Ueda, Rev. Mod. Phys. {\bf 63}, 239 (1991).

\end{thebibliography}
\end{document}